# Designing Incoherent Multispectral Diffractive Optical Networks for Satellite Image Classification

## Authors

Xilin Yang[1,2,3], Md Sadman Sakib Rahman[1,2,3], Ryan Lee[1], Aydogan Ozcan[1,2,3,*]

[1]Electrical and Computer Engineering Department, University of California, Los Angeles, CA, 90095, USA.

[2]Bioengineering Department, University of California, Los Angeles, CA, 90095, USA.

[3]California NanoSystems Institute (CNSI), University of California, Los Angeles, CA, 90095, USA.

* ozcan@ucla.edu

## Abstract

Optical neural networks and diffractive optical computing provide a promising route for performing task-specific inference with high parallelism and lightweight electronic post-processing. These advantages are especially attractive for multispectral satellite imaging, where conventional workflows rely on transmitting high-dimensional image cubes to the ground for downstream electronic analysis. Here, we introduce a computationally efficient surrogate model-guided framework for designing incoherent multispectral diffractive optical networks that directly output class probabilities onboard, eliminating the need to capture and transfer the full multispectral data cube for ground-side inference. We first train a physically compatible surrogate model, represented by a non-negative fully connected model that accommodates a generalized detector readout, yielding a task-specific non-negative multispectral intensity transformation that can be transferred to an optical processor. We then synthesize its diffractive counterpart through data-free point-spread-function-based optimization under spatially incoherent illumination. Using this approach, we numerically demonstrate a six-spectral-channel diffractive classifier for tri-class satellite-scene recognition, achieving a blind-testing accuracy of 79.8% for artificial surface, water body, and vegetated land images, closely approaching the 81.8% accuracy of its surrogate. We further demonstrate a more compact three-spectral-channel design that numerically achieves 72.7% blind-testing accuracy, compared with 75.1% for its corresponding surrogate model. This surrogate model-guided strategy provides a practical pathway for designing compact, task-specific incoherent multispectral optical processors for resource-constrained remote-sensing inference.

## Introduction

Optical computing offers a compelling route to reduce the electronic burden of information processing by exploiting the parallelism of light propagation[1–7]. Diffractive optical networks, in particular, can transform optical inputs through passive structured layers and generate task-specific outputs on the detector plane[8–20]. This capability is attractive for remote sensing, where multispectral imaging provides rich spatial and spectral information for Earth monitoring[21–23]. However, conventional satellite workflows typically require high-dimensional multispectral image cubes to be captured, transferred to the ground, and processed using external computing resources. This data-transfer-and-inference pipeline introduces bandwidth, latency, and hardware burdens, motivating compact onboard inference architectures that can process multispectral scenes directly at the point of acquisition[24–27].

Satellite multispectral scenes are naturally intensity-encoded and are more appropriately modeled under spatially incoherent illumination, while classification of such scenes with optical processing requires the optical processor to jointly handle multiple wavelength channels. These requirements make the design of incoherent multispectral diffractive networks substantially more challenging than that of coherent or single-wavelength optical classifiers[28]. In addition, direct end-to-end optimization under physically accurate incoherent propagation can be computationally expensive[28], limiting systematic exploration of network architectures, hyperparameter sweeps, wavelength combinations, and detector-plane designs. A surrogate model-based framework provides a practical way to address this challenge by first constructing a physically compatible computational surrogate and then transferring its learned transformation into an optical processor.

Here, we introduce an efficient framework for designing incoherent multispectral diffractive optical networks for satellite image classification. Once optimized, the designed system can directly output class probabilities from the monitored field of view onboard, without capturing and transferring the full multispectral data cube to external computing resources for inference. To design the system, we first train a non-negative fully connected surrogate model with a generalized detector readout to learn a task-specific, intensity-compatible multispectral transformation. This surrogate model then serves as the target for data-free point-spread-function-based optimization of a multi-layer incoherent diffractive processor. Using this framework, we numerically demonstrate a six-spectral-channel diffractive classifier for tri-class satellite-scene recognition, achieving a blind-testing accuracy of 79.8% for images of artificial surface, water body, and vegetated land, closely approaching the 81.8% accuracy of its surrogate. We further demonstrate a compact three-spectral-channel design that numerically achieves 72.7% blind-testing accuracy, compared with 75.1% for its corresponding surrogate model, highlighting the flexibility of this strategy for spectrally efficient onboard optical inference.

Figure 1 illustrates the overall concept of our work for classifying multispectral satellite scenes. In the conventional remote-sensing workflow (Fig. 1a–c), a monitoring satellite captures a multispectral image cube through a telescopic imaging system and a multispectral sensor, after which the resulting high-dimensional data are downlinked to the ground for inference using external computing resources and large-scale electronic neural networks. To eliminate the need for capturing and transferring large amounts of data to the ground, we conceptualize an onboard incoherent multispectral diffractive optical classifier (Fig. 1d), in which the multispectral optical information is processed by a compact diffractive network and a detector plane to directly output class probabilities with minimal electronic post-processing.

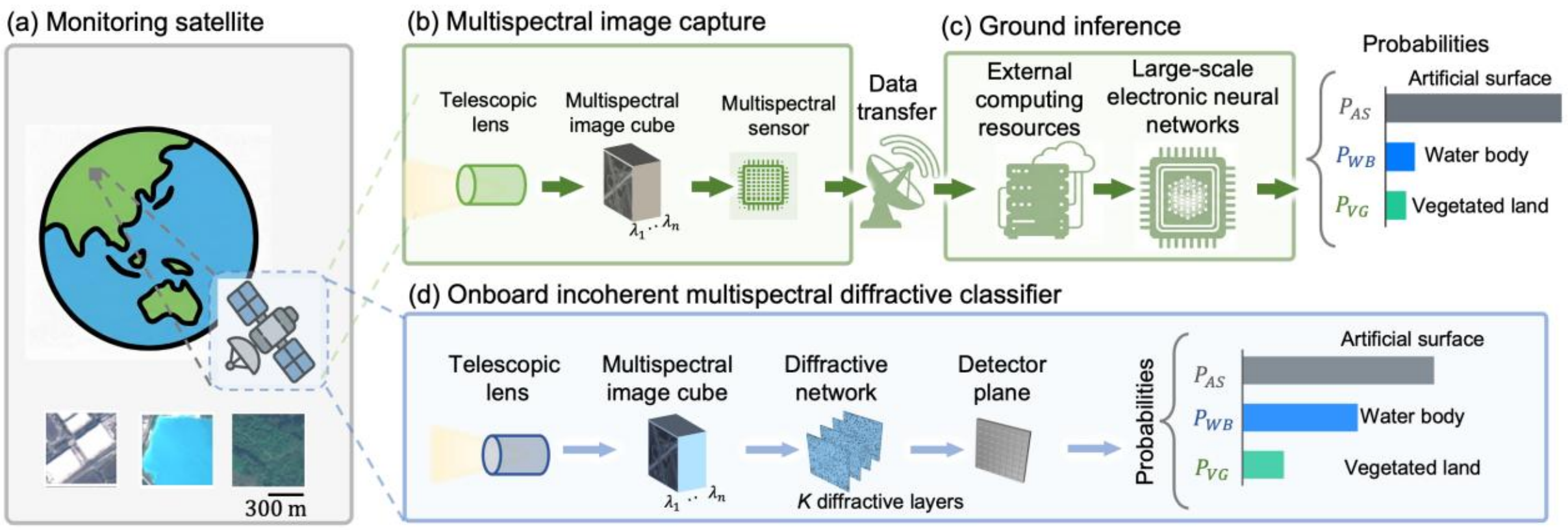

**Figure 1. Onboard multispectral diffractive classification for satellite imaging.** (a) Monitoring satellite. (b) Acquisition of a multispectral image cube using a telescopic lens and a multispectral sensor. (c) Conventional ground-side inference, where the captured data are transferred to external computing resources and processed by electronic neural networks. (d) Onboard diffractive optical classifier, where the multispectral image is processed by a diffractive network and detector plane to output class probabilities.

# Results

## Surrogate model for incoherent multispectral diffractive optical networks

To efficiently design an incoherent multispectral diffractive optical network for satellite image classification, we first establish a surrogate model that serves as the computational blueprint for the optical system. This surrogate model is designed to satisfy two requirements: it should achieve strong tri-class classification performance on the EuroSAT satellite-image dataset[29], and it should yield a non-negative linear mapping that can be physically transferred to an intensity-only incoherent diffractive optical processor. Fig. 2 summarizes this overall design and inference workflow. Figs. 2a–c illustrate the digital and optical design stages, while Fig. 2d shows the final inference process of the optimized optical classifier.

As shown in Fig. 2a, we first perform in silico training of a non-negative fully connected (NNFC) model using data-driven optimization with a cross-entropy objective. The satellite images are grouped into three target classes: (i) artificial surface, (ii) water body, and (iii) vegetated land. The NNFC model maps the input multispectral image cube into a compact set of effective output sensor pixel readouts, where a pair of pixel groups is dedicated to each semantic class. The detector-plane design is illustrated in Fig. 2c. Each class is assigned a spatially multiplexed detector pair, and the corresponding intensity readouts are used to construct class scores and predicted probabilities. To improve the expressiveness of this compact detector readout, we introduce a generalized detector formulation in which each pair of intensity measurements undergoes a lightweight digital transformation parameterized as a learned rational function. These generalized detectors are trained jointly with the NNFC model, allowing the surrogate model to learn both a physically transferable non-negative linear mapping and a compact decision rule tailored to the detector-plane architecture. The detailed mathematical formulation of the generalized detector readout is provided in the Methods section.

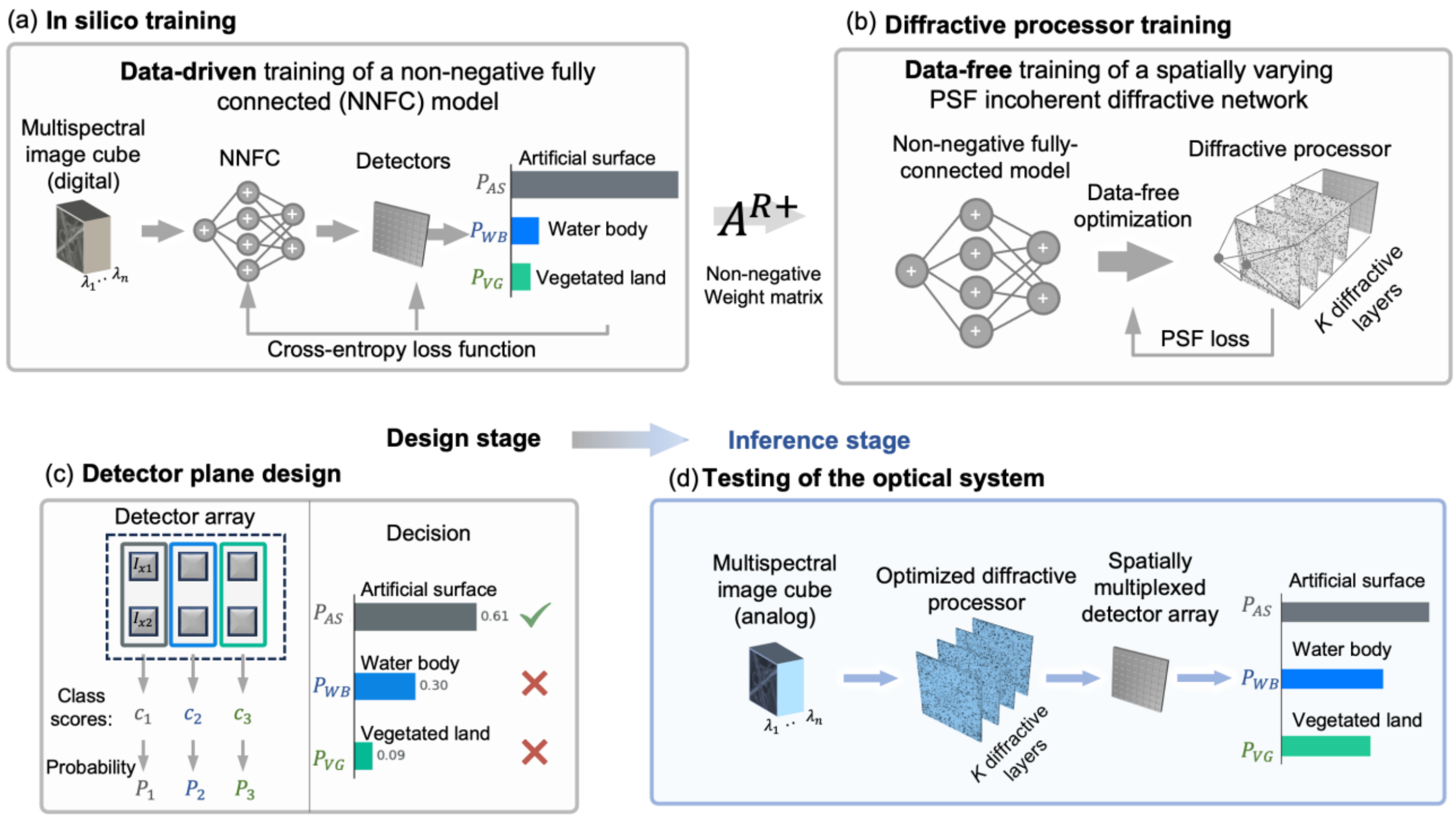

**Figure 2. Surrogate model-guided design of the multispectral diffractive optical classifier.** (a) In silico training of a non-negative fully connected surrogate model using labeled multispectral satellite images. (b) Data-free point-spread-function-based optimization of the incoherent diffractive processor to approximate the learned linear transformation. (c) Detector-plane design and generalized detector readout for generating class scores and probabilities. (d) Inference with the optimized diffractive optical system.

Once trained, the NNFC surrogate model provides the target non-negative intensity transformation to be implemented by the incoherent diffractive optical processor. As illustrated in Fig. 2b, this learned linear mapping is subsequently transferred into an incoherent diffractive network design through data-free point-spread function (PSF)-based optimization. The resulting optical system, together with the generalized detector readout, forms the final multispectral diffractive classifier, as shown in Fig. 2d and reported in the following sections.

## Designing incoherent multispectral diffractive optical networks

Once the surrogate model is trained, we use its learned non-negative linear transformation as the target spatially and spectrally varying PSFs for designing the corresponding incoherent multispectral diffractive optical network. As illustrated in Fig. 2b, this optical design stage is carried out using a data-free PSF-based optimization strategy, where the diffractive processor is optimized to approximate the surrogate model's target transformation through an incoherent propagation model. Rather than directly training the optical network using labeled multispectral images, we sequentially illuminate the input aperture with single-pixel excitations and optimize the resulting spatially and spectrally varying incoherent point-spread functions to reproduce the desired detector-plane intensity mapping[11,28,30]. This strategy avoids the substantial computational burden associated with statistical averaging over a large number of random phase functions at every training step, while preserving a physically meaningful connection between the digital target and the diffractive optical processor.

After training, the optimized diffractive processor is combined with the spatially multiplexed detector array and generalized detector readout for end-to-end inference, as shown in Fig. 2d. The multispectral optical information from the input scene is first propagated through the diffractive layers, producing wavelength-dependent intensity distributions at the detector plane. The wavelength-selective detector-region readings are then summed across the spectral channels and transformed into final class scores and probabilities using the same generalized detector readout learned during the surrogate model training stage. In this way, the designed optical system is optimized to implement the main linear transformation of the hybrid classifier, while only lightweight digital operations are retained after optical detection.

We first numerically evaluate a six-spectral-channel incoherent multispectral diffractive classifier. The blind-testing results of the corresponding surrogate model and its optical counterpart are summarized in Fig. 3. As shown in Fig. 3a, the surrogate model numerically achieves a blind-testing accuracy of 81.8% for satellite images of artificial surface, water body, and vegetated land, with one-vs-rest AUC values of 0.94, 0.97, and 0.93 for artificial surface, water body, and vegetated land classes, respectively. The data-free optimized diffractive processor numerically achieves a comparable blind-testing accuracy of 79.8% (Fig. 3b), corresponding to only a 2.0 percentage-point reduction relative to its surrogate. These results confirm that the PSF-based data-free optimization successfully transfers the classification-relevant transformation learned by the surrogate model into a physically realizable incoherent diffractive optical processor.

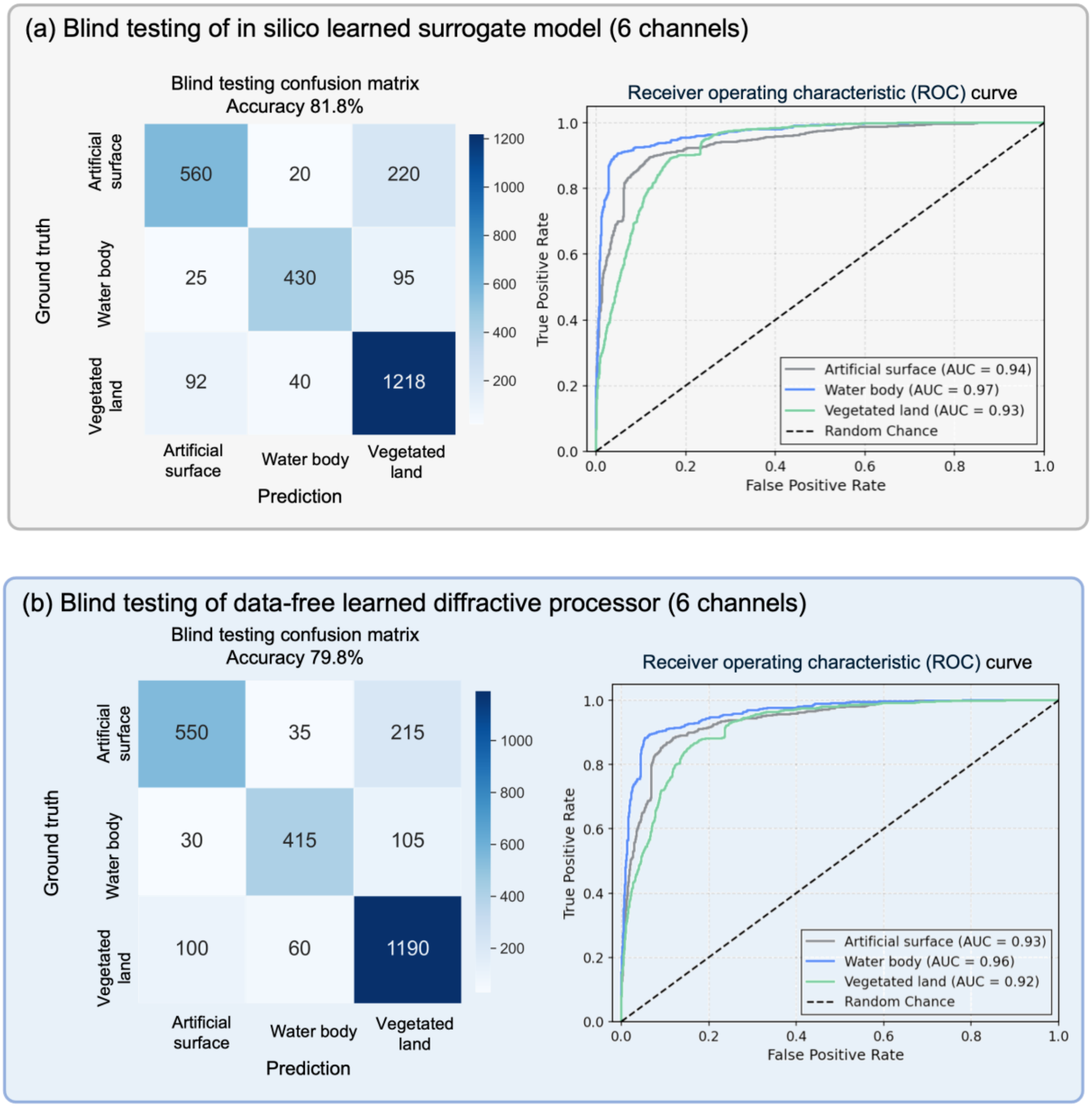

**Figure 3. Blind-testing performance of the six-spectral-channel digital and optical classifiers.** The numerical classification results are shown. (a) Confusion matrix and receiver operating characteristic curves of the six-spectral-channel surrogate model, achieving 81.8% blind-testing accuracy. (b) Corresponding numerical results of the six-spectral-channel diffractive optical processor, achieving 79.8% blind-testing accuracy.

The confusion matrices in Fig. 3 further show that the diffractive optical network preserves the surrogate model's overall class-discrimination behavior. In both models, the strongest classification performance is observed for vegetated land and water body, while a larger fraction of errors involves confusion between artificial-surface scenes and the other two classes. Importantly, the transition from the in silico model to the diffractive optical design introduces only modest changes in the confusion matrix pattern, suggesting that the diffractive network accurately retains the principal decision structure of the surrogate rather than merely matching the aggregate accuracy.

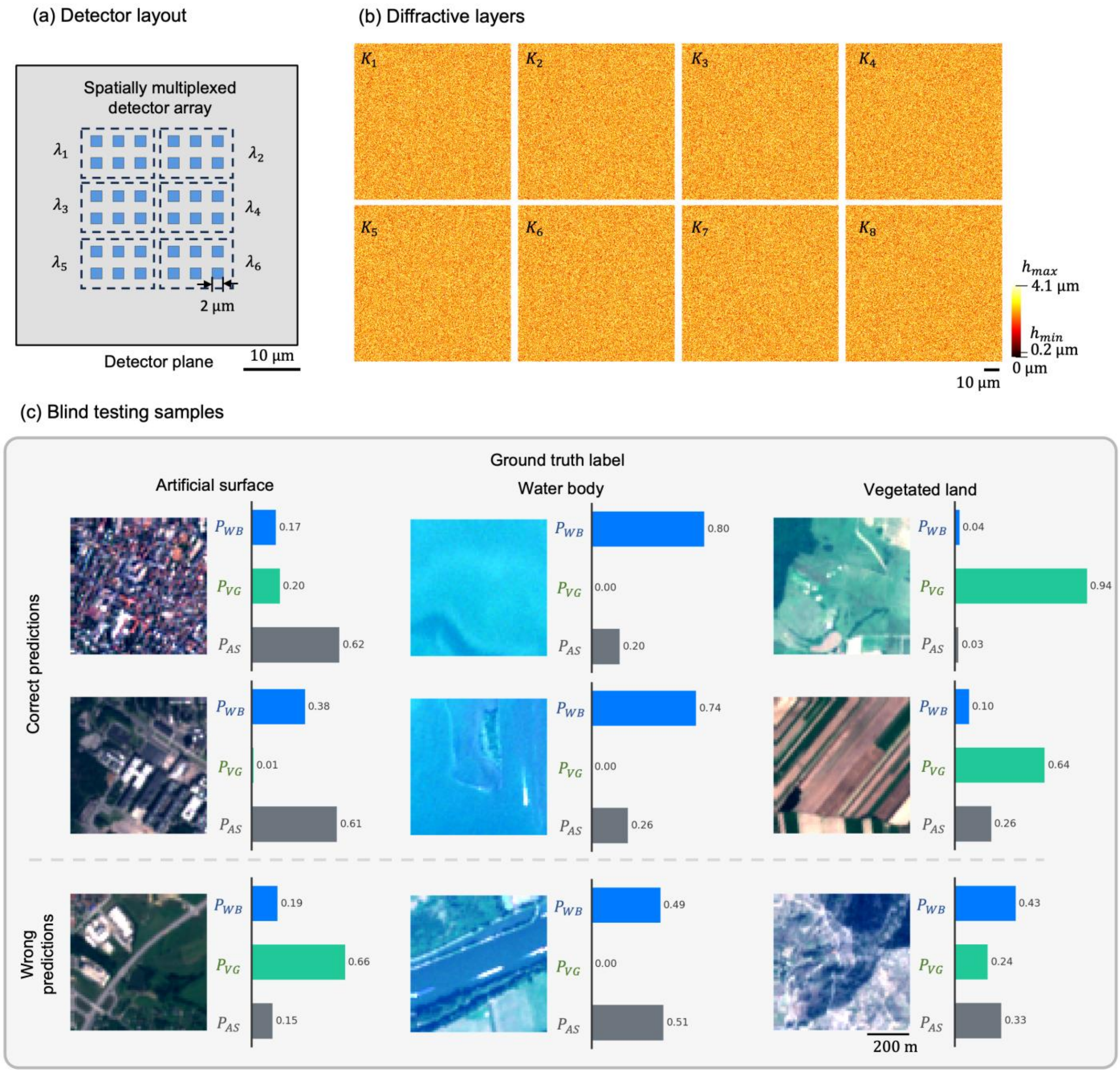

**Figure 4. Six-spectral-channel diffractive optical classifier and representative predictions.** The optical design is visualized. (a) Detector-plane layout used in the six-channel system. (b) Optimized height maps of the diffractive layers. (c) Representative blind-testing samples with ground-truth labels and predicted class probabilities, including correct and incorrect classifications.

Figure 4 provides further details on the six-spectral-channel optical classifier and representative blind-testing examples. The detector-plane design is shown in Fig. 4a, where the spatially multiplexed detector array accommodates the detector regions assigned to the six selected wavelength channels. Figure 4b shows the optimized height profiles of the diffractive layers used in the optical processor. Representative blind-testing samples are presented in Fig. 4c, together with the corresponding class-probability outputs for artificial surface, water body, and vegetated land classes. The correctly classified examples illustrate that the optical system produces strongly class-selective probability responses across visually distinct land-cover types. The misclassified examples further reveal cases in which the inferred probabilities become less separated or shift toward competing classes, consistent with the residual errors observed in the confusion matrix. Together, Figs. 3 and 4 demonstrate that the surrogate model-guided PSF optimization enables an incoherent multispectral diffractive optical network to perform tri-class satellite image classification with its numerical performance closely approaching that of its surrogate model.

We next examined how the surrogate model-based framework can guide more compact system designs by reducing the number of spectral channels. We used the surrogate model to select the wavelengths for the six- and three-spectral channel designs (see Methods for details). Specifically, we designed a compact three-channel incoherent multispectral diffractive classifier, as shown in Fig. 5. The corresponding surrogate model achieves a blind-testing accuracy of 75.1% (Fig. 5c), while the data-free optimized diffractive processor reaches 72.7% (Fig. 5d), leaving a 2.4 percentage-point gap between the digital and optical implementations. These numerical results show that the demonstrated design strategy remains effective even under substantially reduced spectral dimensionality, enabling a more compact optical classifier while preserving meaningful tri-class performance.

## Discussion

In this work, we introduced a surrogate model-guided framework for designing incoherent multispectral diffractive optical networks for satellite image classification. Instead of directly optimizing a diffractive processor from labeled image data, we first trained a physically compatible surrogate model that learns a non-negative, fully connected intensity mapping and a compact, generalized detector readout for the target classification task. The learned transformation was then transferred to its optical counterpart through data-free PSF-based optimization of a multi-layer incoherent diffractive processor. Using this strategy, we designed multispectral diffractive classifiers that closely approach the performance of their corresponding surrogate models, including a compact design that uses only three spectral channels while retaining meaningful classification performance.

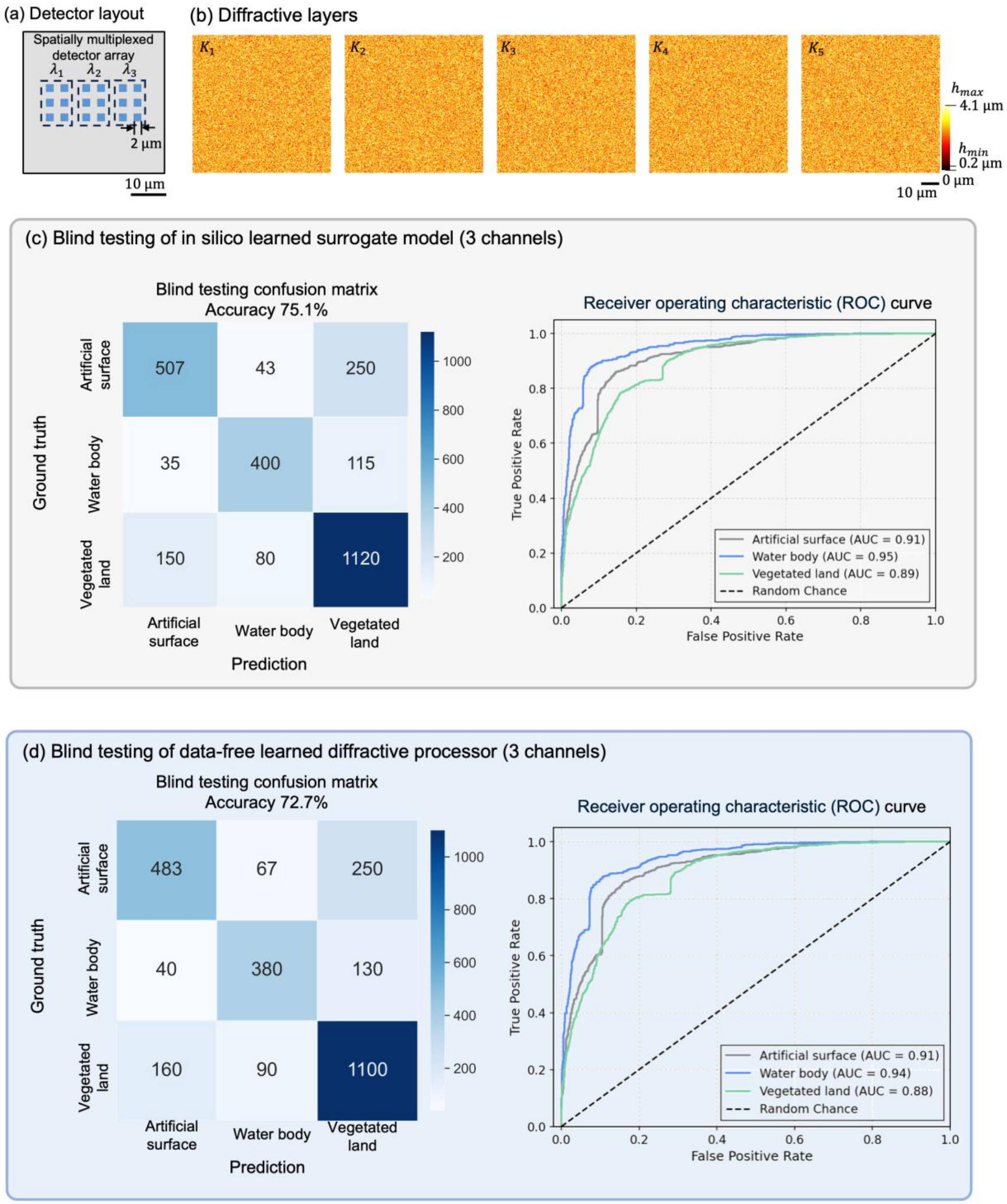

**Figure 5. Compact three-spectral-channel diffractive optical processor.** (a) Detector-plane layout of the three-channel system. (b) Optimized height maps of the five diffractive layers. (c) Blind-testing confusion matrix and receiver operating characteristic curves of the three-channel surrogate model, numerically achieving 75.1% accuracy. (d) Corresponding numerical results of the data-free optimized diffractive optical processor, achieving 72.7% accuracy.

A central contribution of this work is the use of a surrogate model as a practical design interface between task-driven learning and physically constrained diffractive optical synthesis. In this framework, the surrogate model is not merely a numerical reference model, but a computationally efficient tool that enables rapid exploration of optical system design choices before committing to computationally expensive diffractive optimization. For example, wavelength-selection strategies, detector readout formulations, and compact architectural configurations can all be examined first in silico using the surrogate model. This "Gedanken design" approach[31] provides a flexible path for reasoning about physically realizable optical processors: one first asks what class of non-negative mappings can solve the task under the desired readout constraints, and then seeks to synthesize a diffractive optical system that implements that mapping. Such a strategy can be especially valuable when optical training is computationally intensive or when the available design space includes many coupled choices across spectral, detector, and diffractive degrees of freedom.

The PSF-based, data-free optimization strategy used here is also critical for making this design pipeline computationally tractable. When spatially incoherent propagation is modeled by averaging over randomized phase realizations, direct task-driven optimization requires repeated field propagation steps, which become costly for large input apertures and multispectral operation. In contrast, the PSF-based strategy exploits the linearity of incoherent intensity transformations and optimizes the response of the system to point-like input excitations, thereby matching the learned target transformation without requiring labeled image samples during optical training. This offers a computationally efficient and modular route for transferring digital mappings into optical hardware. At the same time, the ability of a diffractive processor to approximate arbitrary multiwavelength intensity transformations remains fundamentally tied to its available degrees of freedom[4,32–34]. In this context, the required number of trainable diffractive features scales as $N \sim 2N_w N_i N_o$ for wavelength-multiplexed linear intensity transformations, where $N_w$, $N_i$, and $N_o$ denote the numbers of spectral channels, input pixels, and output pixels, respectively. This relationship highlights the importance of compact task formulations, reduced spectral budgets, and detector-plane co-design when scaling incoherent diffractive classifiers to more complex remote-sensing problems.

Several limitations and future directions remain. First, the present study focuses on numerical design and blind testing under simulated spatially incoherent illumination. Experimental implementations will require further consideration of fabrication accuracy, alignment tolerance, material absorption, detector noise, and the stability of the optical system under practical satellite operating conditions. Diffractive processors that are resilient against such experimental nonidealities can be designed by incorporating these nonidealities into the training simulations[35]. While the current model assumes spatially incoherent illumination, real remote-sensing scenarios may involve partially coherent or mixed-coherence conditions depending on the scene, imaging geometry, and optical architecture. Prior work has demonstrated that diffractive processors can operate under partially coherent illumination[30]. This provides a route to extending our classifier designs to partially coherent inputs, which we leave as future work. Finally, the present system implements a non-negative linear transformation optically and retains a lightweight nonlinear readout digitally through generalized detectors. Future work could explore richer physically compatible non-negative functions, hybrid cascaded optical–digital inference schemes, or alternative detector-plane formulations that further increase decision capacity while preserving the low-power and compact nature of diffractive optical processors. Together, these directions point toward a broader class of digitally designed incoherent optical systems for onboard remote sensing, resource-constrained classification, and other multispectral inference tasks.

## Materials and Methods

### Dataset and preprocessing

We used the EuroSAT[29] dataset as the base remote-sensing benchmark and constructed a three-class multispectral classification task by grouping the original EuroSAT categories into water body, artificial surface, and vegetated land images. Specifically, River and SeaLake were grouped as water body; Highway, Industrial, and Residential were grouped as artificial surface; and all remaining categories were grouped as vegetated land. The full dataset contains 13 spectral bands centered at 443, 490, 560, 665, 705, 740, 783, 842, 865, 945, 1375, 1610, and 2190 nm. We used a fixed random seed to generate an 80/10/10 training, validation, and test split, and reused the same split across all digital and optical numerical experiments.

We scaled the training inputs to [0, 1] using a single minimum and maximum computed across all training samples and spectral channels. We used the same values to normalize the validation and test inputs. Each sample was represented as a multispectral image cube with $N_w$ spectral channels, where $N_w$ depends on the specific design under consideration. The main diffractive optical classifier uses six spectral channels centered at 490, 665, 783, 842, 1610, and 2190 nm, while the compact design uses three spectral channels centered at 490, 842, and 2190 nm. Each multispectral image was placed at the input plane of the diffractive optical network as a $64 \times 64$-pixel image with a physical pixel size of 400 nm.

### Designing the surrogate model using a non-negative fully connected model

To obtain a compact and physically compatible surrogate model, we trained a non-negative fully connected classifier using a constrained linear model. Let the multispectral input be $I_{in} \in \mathbb{R}_{+}^{H\times W\times C}$. We flatten each spectral channel into a non-negative vector $x_i \in \mathbb{R}_{+}^{N_i}$, where H, W, and C denote the image height, width, and number of spectral channels, respectively, with C=$N_w$, and $N_i = H \times W$ is the number of useful pixels at the input aperture. We denote the bias-adjusted detector readings by $y \in \mathbb{R}^{N_o}$, where $N_o$=2$N_{class}$ and $N_{class}$ is the total number of classes (three unless otherwise stated). The NNFC model applies a grouped linear transformation across the spectral channels with a non-negativity constraint on the weights:

$$y = \sum_{i=1}^{N_w} A(W_i)x_i + b$$

where $W_i$ and $b$ denote the weight matrix associated with the i-th spectral channel and the bias term, respectively, and $A(\cdot)$ is a constraint function used to enforce non-negative weights. Only the non-negative linear component, corresponding to the first term, was implemented by the diffractive optical processor. The learned bias was unconstrained and was applied digitally as a fixed offset to the non-negative optical readings before calculation of the class scores. We considered $A(\cdot) \in \{\mathrm{ReLU}(\cdot), \mathrm{SoftPlus}(\cdot)\}$, where:

$$\mathrm{ReLU}(x) = \max(x, 0)$$

$$\mathrm{SoftPlus(x)} = \log(1 + \exp(x))$$

The subscript $i$ denotes the corresponding spectral channel, and the weights are initialized from a uniform distribution $U[10^{-4}, 1]$.

The resulting detector outputs are converted into class scores $C_n$ using the generalized detector readout described in the following subsection. These class scores are converted into class probabilities using the softmax function:

$$P_n = \frac{\exp(C_n)}{\sum_{m=1}^{N_{\mathrm{class}}} \exp(C_m)}$$

where $P_n$ denotes the predicted probability of the n-th class. The surrogate model is trained using the multiclass cross-entropy loss:

$$\mathcal{L}_{\mathrm{CE}} = -\sum_{n=1}^{N_{\mathrm{class}}} t_n \log(P_n)$$

where $t_n$ is the one-hot encoded ground-truth label for class n.

We used the Adam[36] optimizer for up to 200 epochs and performed a grid search over batch size, learning rate, and the non-negative activation constraint. The model from the epoch with the highest validation accuracy was selected for subsequent optical design. The best-performing configuration used a batch size of 64, ReLU[37] activation, and a learning rate of $1 \times 10^{-4}$. The six-channel wavelength set was obtained by successive exhaustive searches over removals of two, two, two, and one spectral channels from the original 13-band set. For the three-channel design, we exhaustively evaluated two-channel combinations, fixed the pair with the lowest validation loss, and evaluated the addition of each remaining channel. Wavelength combinations were selected based on validation loss. Training and testing were performed on a workstation equipped with an NVIDIA GeForce RTX 3090 Ti graphics processing unit, 256 GB of random-access memory, and an Intel Core i9 central processing unit. The surrogate model and diffractive-network simulations were implemented using Python 3.9 and PyTorch[38] 2.1.0 with CUDA 12.1.

**Generalized detector readout**

Building on the differential detector design[39], each class $n$ is assigned two detector outputs, $y_{n,1}$ and $y_{n,2}$. A differential detector pair computes the corresponding class score as: $C_n = \frac{y_{n,1}-y_{n,2}}{y_{n,1}+y_{n,2}}$. To increase the expressiveness of the detector-plane readout while retaining lightweight post-detection computation, we generalize this formulation to a rational function of the paired detector outputs:

$$C_n = \frac{\sum_{p=0}^{d}\sum_{q=0}^{d} a_{n,p,q} y_{n,1}^{p} y_{n,2}^{q}}{\sum_{p=0}^{d}\sum_{q=0}^{d} b_{n,p,q} y_{n,1}^{p} y_{n,2}^{q}}$$

where $C_n$ is the class score for the $n$-th class, and $a_{n,p,q}$ and $b_{n,p,q}$ are trainable coefficients. The parameter $d$ is set to 2 unless otherwise stated. This generalized formulation relaxes the fixed functional form of the conventional differential detector pair, and all coefficients are jointly optimized with the NNFC parameters using the same optimizer. The conventional differential detector can be recovered as a special case with $d = 1$ by setting $a_{n,1,0} = b_{n,1,0} = b_{n,0,1} = 1$ and $a_{n,0,1} = -1$, with all remaining coefficients set to zero. The generalized readout was initialized to the conventional differential form, and FP32 machine epsilon ($\varepsilon \approx 1.19 \times 10^{-7}$) was added to the denominator in both readout formulations. No numerical instability was observed during training. For the differential-detector baseline, all other model and training settings can be kept unchanged, while the readout coefficients can be fixed to the conventional differential form rather than being jointly optimized.

**Designing incoherent multispectral diffractive optical networks**

The diffractive network consists of multiple phase-only diffractive layers separated by free-space propagation distances and is optimized in silico using a scalar diffraction model. Unless otherwise stated, the main multispectral design contains eight diffractive layers with $512 \times 512$ diffractive features per layer, while the compact three-spectral-channel design contains five diffractive layers with $384 \times 384$ diffractive features per layer. Both designs used a diffractive-feature pitch of 200 nm, with 10 μm separations between the input plane and the first layer, between consecutive diffractive layers, and between the final layer and the detector plane.

We first describe the coherent optical propagation model at a given wavelength $\lambda$, which forms the basis for modeling spatially incoherent propagation. For a multispectral scene, the same propagation model is used for different wavelengths, and the resulting intensities over the wavelength channels are integrated. Each selected band was modeled at its center wavelength, without integration over the finite spectral passband. All selected wavelength contributions were integrated with equal weights, without wavelength-dependent gain calibration. In the numerical readout, only the contribution at the assigned wavelength was retained in each wavelength-specific detector region, corresponding to ideal bandpass selection. Each detector region produced one reading by integrating the retained optical intensity over its area. The retained contributions were summed by class and detector index, then scaled using the common normalization factor from the final training iteration of each design. We then added the learned bias and evaluated the class scores. A filter-free design could instead penalize off-band power in wavelength-assigned detector regions during PSF optimization to suppress spectral cross-talk. The selected wavelengths are indexed from shortest to longest as $\lambda_1, \ldots, \lambda_{N_w}$. The particular spatial assignment is not a fundamental restriction on the target transformation, since diffractive processors can be trained to implement the corresponding output permutations[40].

Each diffractive layer modulates the incident optical field through a spatially varying phase delay:

$$t^k(x, y) = \exp\{j\phi^k(x, y)\}$$

where $\phi^k(x, y)$ denotes the phase modulation introduced by the $k$-th diffractive layer. Free-space propagation between two consecutive planes is modeled using the angular spectrum approach[1]:

$$u(x, y, z + d_k) = \mathcal{F}^{-1}\{\mathcal{F}\{u(x, y, z)\} \cdot H(f_x, f_y; d_k)\}$$

where $u(x, y, z)$ denotes the optical field before propagation and $u(x, y, z + d_k)$ is the field after propagating a distance $d_k$. $\mathcal{F}$ and $\mathcal{F}^{-1}$ denote the two-dimensional Fourier transform and its inverse, respectively, while $f_x$ and $f_y$ denote the spatial frequencies. The corresponding free-space transfer function is given by:

$$H(f_x, f_y; d_k) = \begin{cases} \exp\left\{j\dfrac{2\pi}{\lambda} d_k \sqrt{1 - (\lambda f_x)^2 - (\lambda f_y)^2}\right\}, & f_x^2 + f_y^2 < \dfrac{1}{\lambda^2} \\ 0, & f_x^2 + f_y^2 \geq \dfrac{1}{\lambda^2} \end{cases}$$

where $\lambda$ is the illumination wavelength. During numerical propagation, the computational windows are padded to twice their original lateral dimensions to improve propagation accuracy.

To model spatially incoherent illumination, we use the direct formulation[28] in which the output intensity is estimated by averaging the squared magnitude of the coherently propagated optical field over 20,000 independent random-phase realizations. For each realization, the phase at each point of the 200 nm input-plane simulation grid is independently sampled from a uniform distribution over $[-\pi, \pi]$. This direct incoherent propagation model was used for all blind-testing evaluations of the diffractive optical networks.

However, to train the optical processors, we employ a data-free, point-spread-function-based optimization strategy[28] under the assumption that light originating from different input pixels is mutually incoherent. The target non-negative linear transformation is first obtained from the trained surrogate model. Each input pixel is illuminated individually and represented by multiple point sources in its PSF forward pass, and the resulting point-spread function is optimized to approximate the corresponding target output-intensity distribution. By iterating over all input pixels, the diffractive layers are optimized to collectively approximate the target incoherent linear transformation without requiring labeled image samples during optical training. In each epoch, input pixels were sampled at random without replacement and grouped into batches of 16. We evaluated all selected wavelengths jointly within each batch and averaged the per-wavelength PSF losses. As for the per-wavelength PSF loss, we use a mean-squared-error loss with a normalization factor:

$$\mathcal{L}_{PSF,\lambda} = \frac{1}{N_i N_o} \sum_{p=1}^{N_i} \sum_{q=1}^{N_o} \left( L_\lambda[p,q] - \sigma \widehat{L_\lambda}[p,q] \right)^2$$

where a single normalization factor $\sigma$ is shared by all wavelength channels within each training batch and accounts for the global intensity mismatch associated with diffraction efficiency. For the pooled least-squares normalization, it is defined as:

$$\sigma = \frac{\sum_{\lambda,p,q} L_\lambda[p,q] \widehat{L_\lambda}[p,q]}{\sum_{\lambda,p,q} \widehat{L_\lambda}[p,q]^2}$$

The sums are evaluated over the wavelengths and PSF entries included in the current training batch. Here, $L_\lambda$ and $\widehat{L_\lambda}$ denote the target and output optical non-negative linear transformations at wavelength $\lambda$, respectively, while the corresponding indices denote the input and output pixels. The common normalization factor $\sigma$ was recomputed after each parameter update using the current batch. During backpropagation, $\sigma$ was held fixed, and gradients were not propagated through its calculation.

To account for material dispersion, the diffractive layers are parameterized using their physical height profiles rather than directly optimizing the phase modulation:

$$h_k = h_{min} + \frac{h_{max} - h_{min}}{2} \left( \sin\left( h_k^{latent} \right) + 1 \right)$$

Here, $k$ indexes the diffractive layers, and $h_k^{latent}$ is the trainable latent variable used to compute the physical height $h_k$. The latent variables were initialized by sampling from a standard normal distribution (mean 0, standard deviation 1). This parameterization constrains the diffractive-layer heights to the fabrication range from $h_{min} = 0.2$ µm to $h_{max} = 4.1$ µm. The corresponding wavelength-dependent phase delay is then calculated using the dispersion of N-BK7 glass[41] and its refractive index:

$$\phi_{k,\lambda}(x,y) = \frac{2\pi}{\lambda} (n_\lambda - 1) h_k(x,y)$$

The diffractive optical models are optimized using the AdamW[42] optimizer with a learning rate of $1 \times 10^{-3}$ for up to 20 epochs or until convergence. We selected the optical checkpoint with the lowest PSF loss for blind testing. Training and testing are performed on the same workstation used to train the surrogate model.

## Data availability

The multispectral EuroSAT dataset is publicly available at https://zenodo.org/records/7711810